\documentclass{mem}
\usepackage{natbib}\usepackage{txfonts}\usepackage{balance}
\usepackage{graphicx}
\usepackage[a4paper,breaklinks,dvipdfm]{hyperref}
\idline{75}{282}
\begin{document}
\def\teff{$T\rm_{eff }$}
\def\kms{$\mathrm {km s}^{-1}$}

\title{Nucleosynthesis in massive AGB stars with delayed
superwinds: implications for the abundance
anomalies in Globular Clusters
}

   \subtitle{}

\author{
D. A. \,Garc\'{\i}a-Hern\'andez\inst{1,2} 
\and A. I. \, Karakas\inst{3} \and M. \, Lugaro\inst{4}
          }

  \offprints{D. A. \,Garc\'{\i}a-Hern\'andez}

\institute{
Instituto de Astrof\'{\i}sica de Canarias, Via L\'actea s/n, 38200 La Laguna (Tenerife), Spain; \email{agarcia@iac.es}
\and
Departamento de Astrof\'{\i}sica, Universidad de La Laguna, 38206 La Laguna (Tenerife), Spain
\and 
Research School of Astronomy \& Astrophysics, Mount Stromlo Observatory, Weston Creek ACT 2611, Australia; \email{akarakas@mso.anu.edu.au}
\and 
Monash Centre for Astrophysics, Monash University, Clayton VIC 3800, Australia; \email{maria.lugaro@monash.edu.au}
}

\authorrunning{Garc\'{\i}a-Hern\'andez}

\titlerunning{Delayed superwind AGBs and GC abundance anomalies}

\abstract{
We present nucleosynthesis predictions for massive (5$-$7 M$_{\odot}$) asymptotic
giant branch (AGB) stars of solar metallicity where we delay the onset of the
superwind to pulsation periods of P =700$-$800 days. We found that delaying the
superwind in solar metallicity massive AGB stars results in a larger production
of s-process elements, something that would be also expected at lower
metallicities. These new models and the available observations show that massive
C-O core AGB stars in our Galaxy and in the Magellanic Clouds experience
considerable third dredge-up (TDU). Thus, if massive AGB stars at the
metallicities of the Globular Clusters (GCs) also experience deep TDU, then
these stars would not be good candidates to explain the abundance anomalies
observed in most GCs. However, more massive AGB stars (e.g., near the limit of
C-O core production) or super-AGB stars with O-Ne cores may not experience very
efficient TDU, producing the high He abundances needed to explain the multiple
populations observed in some GCs.
\keywords{Stars: AGB and post-AGB -- Stars: abundances --
Stars: atmospheres -- Stars: evolution -- nuclear reactions, nucleosynthesis,
abundances -- Galaxy: globular clusters }
}
\maketitle{}

\section{Motivation}

Garc\'{\i}a-Hern\'andez et al. (2006, 2007, 2009) identified several Galactic
and Magellanic Cloud massive (M $>$ 4$-$5 M$_{\odot}$) asymptotic giant branch
(AGB) stars within a sample of OH/IR stars. The strong enhancements of the
neutron-rich element Rb found together with the fact that these stars are O-rich
and Li-rich, support the prediction that hot bottom burning (HBB) and an
efficient third dredge-up (TDU) have occurred in these stars. Rb is believed to
be produced by the s-process in AGB stars. A Rb enrichment over the s-process
elements Sr, Y, and Zr is evidence for the efficient operation of the
$^{22}$Ne($\alpha$, n)$^{25}$Mg neutron source in massive AGB stars (e.g.,
Garc\'{\i}a-Hern\'andez et al. 2006). This is because the production of Rb is
sensitive to the high neutron density associated with the $^{22}$Ne neutron
source. These observational results are particularly important because there is
a lack of observational evidence for constraining stellar models of massive
AGBs, and especially the TDU and HBB efficiencies. These two last points are
much debated in the context of the globular cluster (GC) abundance anomalies
(see e.g., Karakas et al. 2006). 

van Raai et al. (2012) studied the Rb production in massive AGB stars at
different metallicities. The qualitative features of the observations could be
reproduced; increasing [Rb/Fe] ratio with increasing stellar mass or decreasing
metallicity. However, the models could not reproduce the most Rb-rich AGB stars.
Possible solutions are: i) to extend the calculations to include models of
masses higher than 6 M$_{\odot}$; ii) to modify the mass-loss rate on the AGB.
Here we present the exploration of these two possible solutions by extending
calculations to masses of up to 9 M$_{\odot}$ and varying the AGB mass-loss rate. We also
briefly discuss the nucleosynthesis results in the context of the abundance
anomalies observed in GCs.

\section{Delayed superwind massive AGB models}

Vassiliadis \& Wood (1993) (VW93 hereafter) noted that there are long period
variables (LPV) stars with periods of 750 days that are probably stars of $\sim$5
M$_{\odot}$ and they recommended a delay to the onset of the superwind in stars
with M $>$ 2.5 M$_{\odot}$. This suggestion is supported by the optical
observations of massive AGBs, which show that the number of dust enshrouded stars
dramatically increases for periods longer than 700 days (Garc\'{\i}a-Hern\'andez
et al. 2007). Here we explore the effect of delaying the onset of the superwind
phase on the AGB nucleosynthesis.

\section{Nucleosynthesis results} 

Stellar evolutionary sequences for massive AGB stars are fed into a post-processing
code in order to obtain nucleosynthesis predictions for elements heavier than iron
(see Karakas et al. 2012 for more details). In short, we consider stellar masses from
5 to 9 M$_{\odot}$ and we compute one evolutionary sequence using the standard VW93
mass-loss prescription. For M = 5$-$7 M$_{\odot}$ models, we compute one evolutionary
sequence using the modified VW93 mass-loss prescription or ``delayed superwind". In
all cases convergence difficulties (i.e., model star had left the AGB once the
envelope mass drops below $\sim$1 M$_{\odot}$) end the calculation after the
cessation of HBB and the effect of these extra thermal pulses (TPs) on the surface
compositions of our model stars is estimated by synthetic evolution (see Karakas et
al. 2012 for more details).

\begin{figure}
\resizebox{\hsize}{!}{\includegraphics[angle=-90,scale=.45,clip=true]{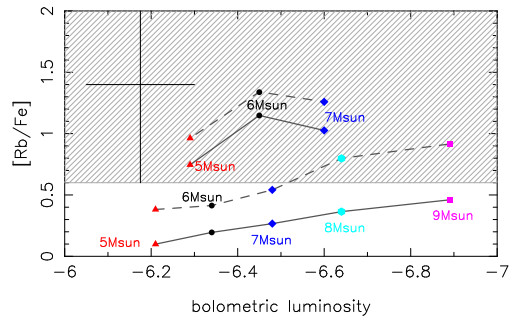}}
\caption{\footnotesize
Bolometric luminosity at the tip of the AGB vs. the [Rb/Fe] abundance from the
last TP (connected by the solid lines) and from the synthetic evolution
calculations (connected by the dashed lines). Models using the Vassiliadis \&
Wood (1993) mass-loss prescription are connected by the lower solid and dashed
lines, and models calculated using a delayed superwind by the upper solid and
dashed lines, respectively. The shaded region indicates the range of observed
[Rb/Fe] (average of 1.4 dex and maximun uncertainty of $\pm$0.8 dex).
}
\label{Fig1}
\end{figure}

In Figure 1 we show the bolometric luminosity at the tip of the AGB versus the
[Rb/Fe] ratios for models calculated using the VW93 mass-loss formula and for models
with a delayed superwind. From Figure 1 we see that models with a delayed superwind
produce higher [Rb/Fe] abundances (shown by the upper solid and dashed lines) and are
found within the shaded region of the observations. This is mostly because these
models experience many more TPs and TDU episodes. The evolution of elements [X/Fe]
lighter and heavier than Fe for the 5 M$_{\odot}$ model with a delayed superwind is
shown in Figure 2, where we see no significant production of elements beyond the
first s-process peak.

\begin{figure}
\resizebox{\hsize}{!}{\includegraphics[angle=-90,scale=.01,clip=true]{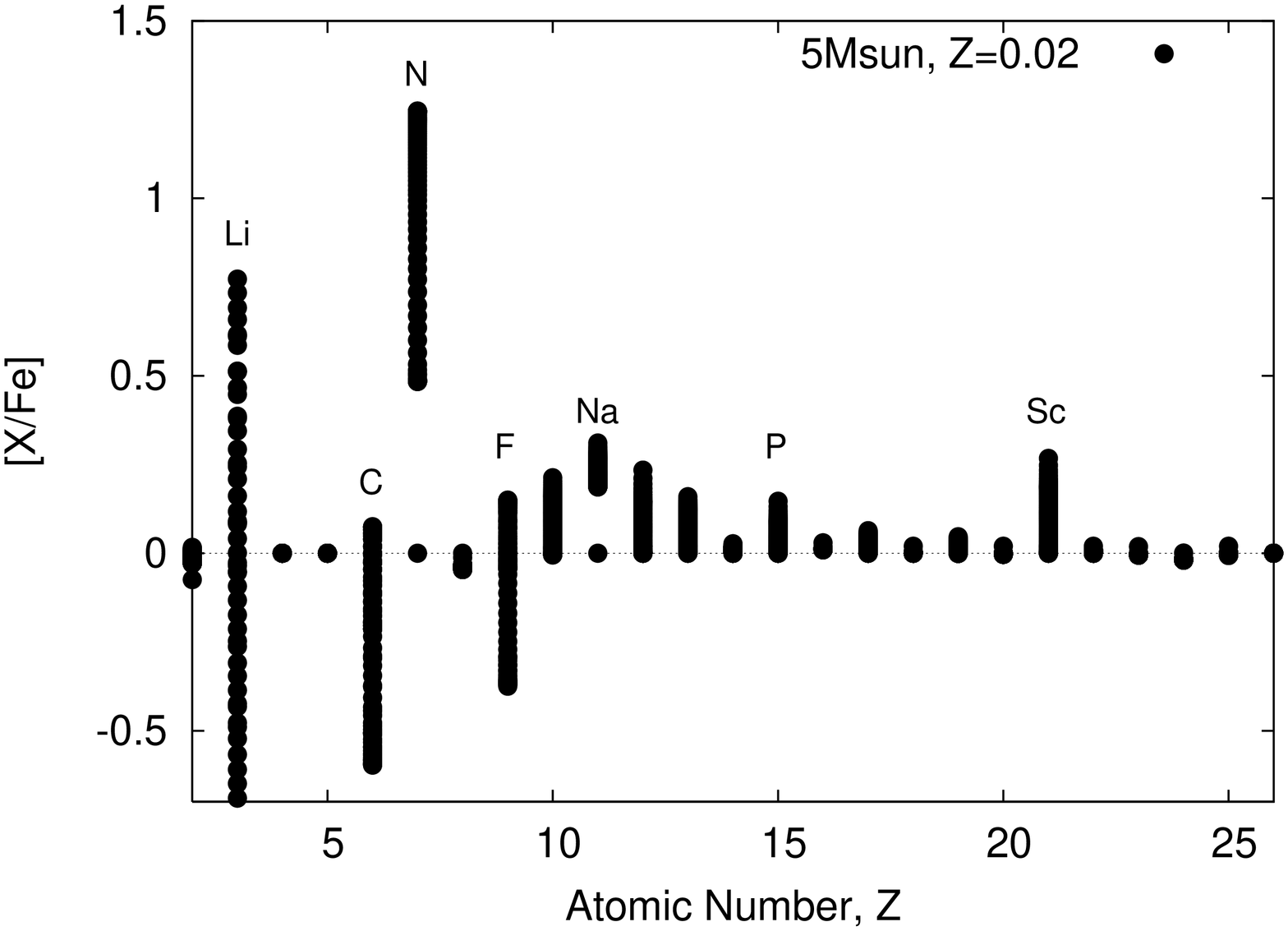}}
\resizebox{\hsize}{!}{\includegraphics[angle=-90,scale=.01, clip=true]{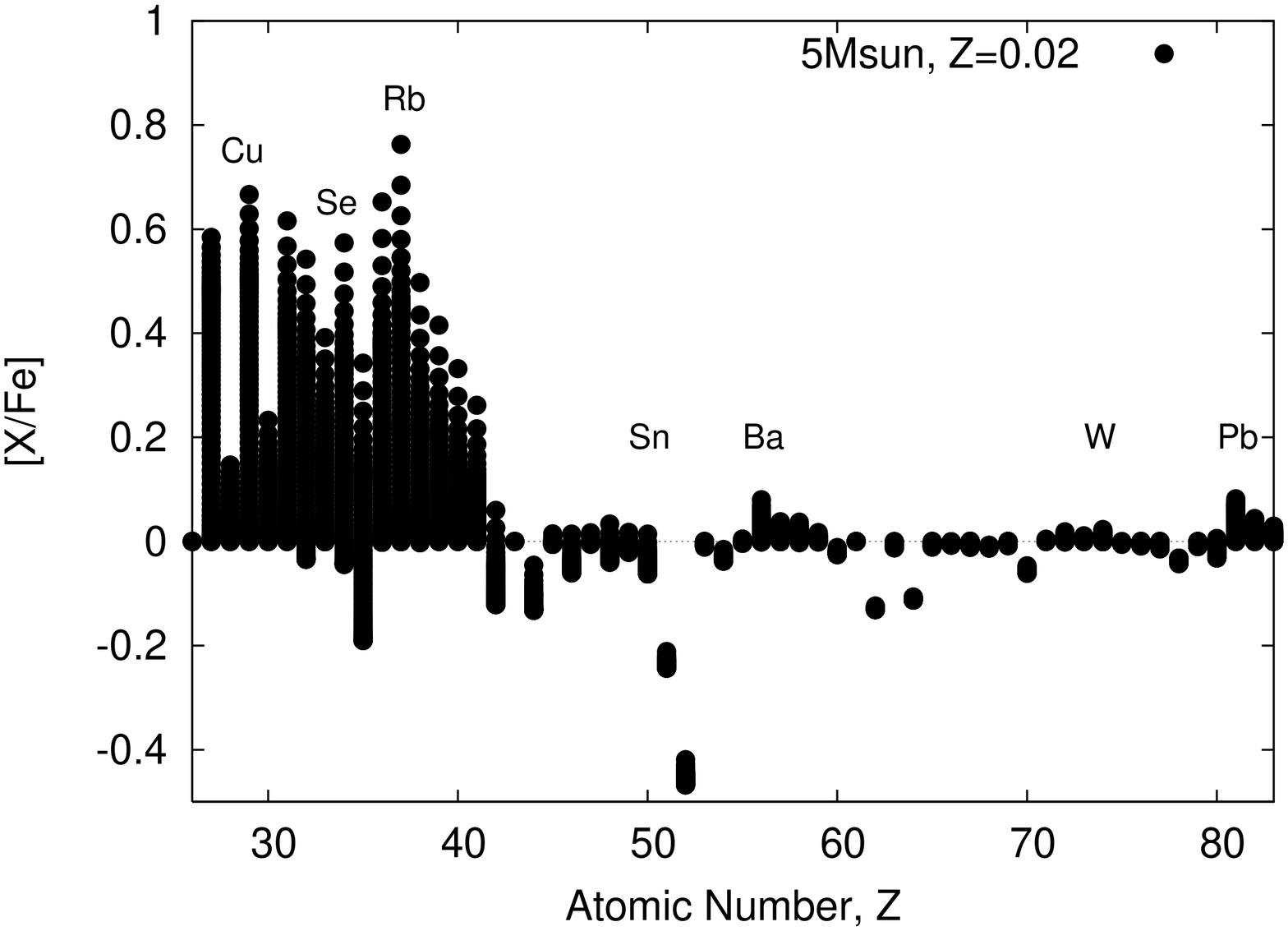}}
\caption{\footnotesize
Evolution of elements [X/Fe] lighter (upper panel) and heavier (lower panel) than Fe
as a function of atomic number, Z, for the 5 M$_{\odot}$ model with a delayed
superwind and nuclear network using 320 species. Included are the approximate
locations of some key elements. Each dot represents the surface composition after a
TP.
}
\label{Fig2}
\end{figure}

\section{Implications for abundance anomalies in GCs}

Delaying the superwind in massive AGB stars results in a far greater production of
neutron-capture elements at solar metallicity and we would expect an extension of
this effect to the lowest metallicities. The new models and the observations by
Garc\'{\i}a-Hern\'andez et al. (2006, 2007, 2009) show that the most massive C-O core
AGB stars of solar metallicity (up to 8 M$_{\odot}$) and at the metallicities of the
Magellanic Clouds experience considerable TDU. Thus, if massive AGBs at the
metallicities of the Galactic GCs (-2.3 $<$ [Fe/H] $<$ -0.7) also experience deep
TDU, then these stars would not be good candidates to explain the abundance anomalies
observed in every well studied GC (e.g., Gratton et al. 2004). 

AGB stars of up to 6$-$7 M$_{\odot}$ might instead be candidates for producing the
neutron-capture elements in some GCs, including M4 (e.g., Yong et al. 2008) and M22
(Roederer et al. 2011). Note that the neutron-capture elements observed in GCs do not
show any correlation with the light-element abundance patterns, ruling out a relation
between the two. However, models of stars with masses near the limit of C-O core
production (here the 8 M$_{\odot}$ model) or super-AGB stars with O-Ne cores may not
experience very efficient TDU together with very hot HBB and depending upon the
model, little pollution from He-intershell material (e.g., Siess 2010; Ventura \&
D'Antona 2011). Thus, our results and theirs show that super-AGB stars can produce
the high He abundances needed to explain the multiple populations observed in the
colour-magnitude (C$-$M) diagrams of some GCs (e.g., Piotto et al. 2005). 

It should be noted, however, that Rb abundances in massive AGBs may be largely
overestimated as a consequence of our imcomplete understanding of their complex
atmospheres (Garc\'{\i}a-Hern\'andez et al. 2009). Model atmospheres of massive AGBs
should be improved (see discussion in van Raai et al. 2012) before reaching a final
conclusion about the connection of these stars and the abundance anomalies in GCs.

\begin{acknowledgements}

D.A.G.H. acknowledges support provided by the Spanish Ministry of Economy and
Competitiveness under grant AYA$-$2011$-$29060. A.I.K. is grateful for the
support of the NCI National Facility at the ANU and the ARC for support through
a Future Fellowship (FT110100475). M.L. thanks the ARC for support through a
Future Fellowship (FT100100305) and Monash University for support through a
Monash Research Fellowship.

\end{acknowledgements}

\bibliographystyle{aa}

\end{document}